\documentclass[prb, preprint, superscriptaddress, endfloats*]{revtex4-1}
\usepackage{amsmath,hyperref,cleveref,amssymb,amsfonts,graphicx,multirow,color,bm,textcomp,mathtools}
\usepackage{paralist}
\usepackage{srcltx}
\usepackage[all,cmtip]{xy}
\usepackage{mathrsfs}
\usepackage{srcltx,soul,subfigure}
\usepackage{threeparttable,dsfont,bbm}
\usepackage{color}
\usepackage{dcolumn}

\newcommand{\bra}[1]{\langle #1 |}
\newcommand{\ket}[1]{| #1 \rangle}

\newcommand{\bee}{\begin{equation}}
\newcommand{\ee}{\end{equation}}
\newcommand{\bma}{\begin{pmatrix}}
\newcommand{\ema}{\end{pmatrix}}
\newcommand{\balig}{\begin{align}}
\newcommand{\ealig}{\end{align}}

\newcommand{\bp}{{\bm p}}
\newcommand{\bb}{{\bm b}}
\newcommand{\bk}{{\bm k}}

\newcommand{\ba}{\begin{eqnarray}}
\newcommand{\ea}{\end{eqnarray}}

\newcommand{\ignore}[1]{}

\begin{document}


\title{Dirac Fermion Cloning, Moir{\'e} Flat Bands and Magic Lattice Constants in Epitaxial Monolayer Graphene} 

\author{Qiangsheng~Lu*}
\affiliation {Department of Physics and Astronomy, University of Missouri, Columbia, Missouri 65211, USA}
\author{Ching-Kai~Chiu*}
\affiliation{Kavli Institute for Theoretical Sciences, University of Chinese Academy of Sciences, Beijing 100190, China} 
\author{Congcong~Le}
\affiliation{Max Planck Institute for Chemical Physics of Solids, 01187 Dresden, Germany}
\author{Jacob~Cook}
\affiliation {Department of Physics and Astronomy, University of Missouri, Columbia, Missouri 65211, USA}
\author{Xiaoqian~Zhang}
\affiliation{Department of Physics and Astronomy, University of Missouri, Columbia, Missouri 65211, USA} 
\author{Xiaoqing~He}
\affiliation {Electron Microscopy Core Facility, University of Missouri, Columbia, Missouri 65211, USA}
\affiliation {Department of Mechanical and Aerospace Engineering, University of Missouri, Columbia, MO 65211, USA}
\author{Mohammad~Zarenia}
\affiliation {Department of Physics and Astronomy, University of Missouri, Columbia, Missouri 65211, USA}
\author{Mitchel~Vaninger}
\affiliation {Department of Physics and Astronomy, University of Missouri, Columbia, Missouri 65211, USA}
\author{Paul F. Miceli}
\affiliation {Department of Physics and Astronomy, University of Missouri, Columbia, Missouri 65211, USA}
\author{Chang~Liu}
\affiliation{Department of Physics, Southern University of Science and Technology, Shenzhen 518055, China} 
\author{Tai-Chang Chiang}
\affiliation {Department of Physics, University of Illinois at Urbana-Champaign, 1110 West Green Street, Urbana, IL 61801-3080, USA }
\affiliation {Frederick Seitz Materials Research Laboratory, University of Illinois at Urbana-Champaign, 104 South Goodwin Avenue, Urbana, IL 61801-2902, USA}
\author{Giovanni~Vignale}
\affiliation {Department of Physics and Astronomy, University of Missouri, Columbia, Missouri 65211, USA}
\author{Guang~Bian}
\affiliation {Department of Physics and Astronomy, University of Missouri, Columbia, Missouri 65211, USA}
\begin{abstract}

{\bf Tuning interactions between Dirac states in graphene has attracted enormous interest because it can modify the electronic spectrum of the two-dimensional material, enhance electron correlations, and give rise to novel condensed-matter phases such as superconductors, Mott insulators, Wigner crystals and quantum anomalous Hall insulators. Previous works predominantly focus on the flat band dispersion of coupled Dirac states from different graphene layers. In this work, we propose a new route to realizing flat band physics in monolayer graphene under a periodic modulation from substrates. We take gaphene/SiC heterostructure as a role model and demonstrate experimentally the substrate modulation leads to Dirac fermion cloning and consequently, the proximity of the two Dirac cones of monolayer graphene in momentum space. Our theoretical modeling captures the cloning mechanism of Dirac states and indicates that flat bands can emerge at certain magic lattice constants of substrate when the period of modulation becomes nearly commensurate with the $\boldsymbol{(\sqrt{3}\times\sqrt{3})R30^{\circ}}$ supercell of graphene. The results show that the epitaxial monolayer graphene is a promising platform for exploring exotic many-body quantum phases arising from interactions between Dirac electrons.}

\end{abstract}


\maketitle 



\section{Introduction}
The discovery of graphene has revolutionized modern condensed matter physics as it provides direct access to the physics of Dirac fermions in solid-state systems\cite{CastroNeto2009, Novoselov2005, Zhang2005}. It also sheds light on the path towards a vast field of novel 2D materials including van der Waals (vdW) materials and topological materials such as quantum spin Hall insulators \cite{CastroNeto2009, Geim2013, Katsnelson2006, RevModPhys.86.959, PhysRevLett.61.2015, RevModPhys.82.3045, RevModPhys.83.1057, Liu2011}. A single layer graphene possesses two copies of Dirac cones residing at the opposite corners of the Brillouin zone, leaving them essentially isolated from each other. Stacking graphene layers in the default order duplicates Dirac cones in the same valley and thus makes the two valleys remain decoupled. Recently, new excitements  in graphene-like systems have arisen as a consequence of the creation of strongly coupled Dirac states in artificially engineered structures such as twisted bilayer graphene (TBG). The interaction between Dirac states from the two layers can be effectively tuned by the angle mismatch and thus leads to emergent collective behaviors of electrons including Mott insulating states, unconventional superconductivity, emergent ferromagnetism, quantum anomalous Hall effects \cite{Bistritzer12233, Ponomarenko2013, Dean2013, Cao20181, Cao20182, Wang2016, Sharpe605, Serlin900, Kaxiras2020}. The essential ingredient for those new emergent states is the nearly dispersionless bands at zero energy in the Moir{\' e} Brillouin zone. Achieving the required large periodicity of Moir\'{e} pattern in real space and the closeness of Dirac cones in momentum space generally requires delicate controls over the twist between two mechanically exfoliated graphene layers, which places stringent constraints on the techniques of sample assembly.  Therefore, there is a pressing need for accessing flat band physics in systems without fine tuning on twist angles. 

Here we report an alternative route to enable interactions between Dirac electrons in a single layer of graphene and realize flat bands with aid from the supporting substrate potential. We observed the cloning of Dirac bands in monolayer graphene epitaxially grown on SiC substates by angle-resolved photoemission spectroscopy (ARPES). The periodic substrate potential brings closer the Dirac states from the two valleys and thus turns on intervalley coupling, which is precisely captured by our tight-binding simulations. Our theory further indicates that the perturbed graphene system with nearly commensurate epitaxial relations hosts absolutely flat bands due to the fact that the AB coupling between the two valleys is absent. The lack of AB coupling yields a chiral symmetry of the low-energy effective Hamiltonian and consequently makes the epitaxial graphene system a natural realization of the chiral-symmetric continuum model proposed by Tarnopolsky {\it et al.}\cite{PhysRevLett.122.106405}. 

\section{Experimental Results}
First, we study the substrate effects on the Dirac states of graphene. Our graphene was grown epitaxially on the Si-face of a 6H-SiC substrate. The lattice structure and epitaxial relation between the graphene overlayer and the SiC substrate are plotted in Fig.~1{\bf a}. The lattice constant of graphene and SiC(0001) surface is 2.46~\AA ~and 3.07~\AA, respectively. The Brillouin zone of graphene and SiC(0001) is depicted by red and blue lines, respectively, in Fig.~1{\bf b}.  ${\rm K}_{\textrm{Gr}}$ and ${\rm K'}_{\textrm{Gr}}$ represent the location of the two valleys of graphene Dirac states. The atomic-resolution STM image of the graphene sample is shown in Fig.~1{\bf c}. A superhexagonal Moir\'{e} pattern with a period $\lambda = 6(1 + \delta)a_\textrm{SiC}$ is observed due to the incommensurate modulation of the SiC interface layer \cite{Conrad2017, PhysRevB.96.035411}. To examine the structural quality, we performed high-resolution TEM measurements on our graphene samples. A typical TEM image is shown in Fig.~1{\bf d}. The sample consists of four well-ordered graphene layers sitting on the carbonized surface of the SiC substrate. A gap between the graphene and the SiC surface is noticeable, indicating a sharp interface between the graphene layers and the substrate. In our experiment, the thickness of graphene layers can be precisely controlled down to a monolayer. We will focus on results obtained from monolayer graphene samples in the following discussion, but the physics discussed here also applies to thicker graphene films.

The fermi surface of the graphene sample mapped by ARPES is presented in Fig.~2{\bf a}. The Brillouin zone of graphene is marked by the blue dashed lines. At the corners of the Brillouin zone, the ${\rm K}_{\textrm{Gr}}$ points, we can see bright fermi surface contours from the Dirac bands of graphene. Note that our sample is $n$-typed doped with fermi level above the Dirac point. Therefore, the fermi surface contours are small circles surrounding ${\rm K}_{\textrm{Gr}}$, as marked by the black arrows. Beside the Dirac states of graphene, there are extra circular contours located inside the Brillouin zone of graphene, which are absent on the fermi surface of freestanding graphene. These new contours marked by blue, green, yellow, and gray arrows are referred to as blue, green, yellow, and gray contours (or cones) in the following discussions. The geometric relations between the Brillouin zones of graphene and SiC is depicted in Fig.~2{\bf b}. The emergent contour with lesser photoemission intensity can be considered as the clones of the graphene Dirac cone generated by the periodic substrate perturbations. Shifting the ${\rm K'}_{\textrm{Gr}}$ by a reciprocal lattice vector of SiC as shown in Fig.~2{\bf c}, we can find the location of the blue and green cones. The yellow and gray contours can be obtained through two successive shifts of the graphene states, see Fig.~2{\bf d}. Taking the substrate interaction as a perturbation, we can attribute the blue (green) and yellow (gray) cones to the first-order and second-order perturbation effects, which also explains that the intensity of yellow (gray) cones is apparently lower than that of blue (green) cones. The locations of all the first-order and second-order clones are summarized in Fig.~2{\bf e}. It is consistent with the experimental observations shown in Fig.~2{\bf a}.

The cloning of Dirac states is deeply rooted in the periodic modulation exerted by the substrate on the electrons in graphene\cite{}. To understand the mechanism of cloning, we performed a tight-binding simulation in which the substrate effect is approximated by a periodic potential acting on the the graphene electrons. The sample is n-type doped, so the fermi level is shifted in the simulation to match the experimental results. There is a good agreement between the experimental and theoretical fermi surfaces as shown in Figs.~3{\bf a} and 3{\bf b}. All first-order and second-order clones in the simulation show up at the locations observed in the experimental results. The spectrum along the line of ‘cut1’ (marked in Fig.~3{\bf a}) is plotted in Figs.~3{\bf c} and 3{\bf d}. The brightness of bands indicates the photoemission intensity in the ARPES spectrum and the spectral weight (the probability of finding the electron with the corresponding energy and momentum) in the TB simulation.  The clones show the same dispersion as the primary Dirac cones but have lesser spectral weight compared to the primary cone, consistent with the perturbative picture. This is further corroborated by the ARPES spectrum taken along the lines of ‘cut2’-‘cut5’. Unlike the primary Dirac cones sitting far apart at the corners of the Brillouin zone, the clones are much closer to each other in momentum space. For example, the distance between the yellow and green clones is 0.21~\AA$^{-1}$ while the spacing between neighboring primary cones is 1.7~\AA$^{-1}$. The smaller distance between clones enables them to overlap in momentum space, see the iso-energy contours in the supplementary information. The yellow and green cones cross at $E = -1.3$~eV while the blue and green cones cross at $E = -1.8$~eV. Such crossings of Dirac states are unavailable in freestanding graphene films. It is worth noting that there is no gap opened at the crossings of the green and blue bands. This is because both green and blue contours originate from the same valley (see Fig.~2{\bf c}) and thus they do not hybridize. Only Dirac bands from different valleys can interact with other other and open hybridization gaps, which will be shown in the following discussions. The cloning of Dirac bands occurs not only in the monolayer graphene but also in thicker films. Figure~3{\bf f} shows the spectra of the primary Dirac cone from samples with different thicknesses, namely, 1~ML, 3~ML, 5~ML. The number of Dirac bands in the spectra indicates the thickness of the sample \cite{Zhou2006, Liu2010}. Despite of the weaker intensity, the clone band (the blue cone marked in Figs.~2{\bf a} and 3{\bf c}) are still observable in the 3-ML and 5-ML samples as shown in Fig.~3{\bf f}. 

\section{Discussions}
The perturbation theory of quantum mechanics can capture the essential physics of Dirac fermion cloning in the graphene/SiC heterostructure. The SiC substrate potential can be treated as a small perturbation and exerts a periodic modulation to the Dirac states of graphene \cite{PhysRevB.96.035411}. Taking $W({\bm x})$ as the SiC potential, the eigenfunction up to the second-order perturbations is given by 
\begin{align}
\ket{\Psi_\bk}=& \ket{\Psi_\bk^0}+  \sum_{\bp\neq \bk} \ket{\Psi_{\bm p}^0} \frac{\bra{\Psi_{\bm p}^0} W \ket{\Psi_{\bm k}^0} }{E_\bk^0 - E_\bp^0} + \sum_{\bp \neq \bk, {\bm l}\neq \bk } \ket{\Psi_{\bm p}^0} \frac{\bra{\Psi_{\bm p}^0} W \ket{\Psi_{\bm l}^0} \bra{\Psi_{\bm l}^0} W \ket{\Psi_{\bm k}^0} }{(E_\bk^0 - E_\bp^0)(E_\bk^0 - E_{\bm l}^0)} \nonumber \\
&- \sum_{\bp\neq \bk} \ket{\Psi_{\bm p}^0} \frac{\bra{\Psi_{\bm k}^0} W \ket{\Psi_{\bm k}^0} \bra{\Psi_{\bm p}^0} W \ket{\Psi_{\bm k}^0}}{(E_\bk^0 - E_\bp^0)^2} -\frac{1}{2} \ket{\Psi_{\bm k}^0} \sum_{\bp\neq \bk} \frac{| \bra{\Psi_{\bm p}^0} W \ket{\Psi_{\bm k}^0}|^2 }{(E_\bk^0 - E_\bp^0)^2}, 
\end{align}		
where `0' indicates the original wavefunction of the graphene in the absence of the SiC potential.  The second term in Eq.~(1) is the first-order correction. To have non-vanishing coefficient $\bra{\Psi_{\bm p}^0} W \ket{\Psi_{\bm k}^0}$, the difference between  $\bp $ and $\bk $ must align with the period of $W({\bm x})$. We note that we simulate the SiC potential in terms of the simplest harmonic form, that is, $W({\bm x})=w\big(\cos({\bm{ b_{1}\cdot x}}) +\cos({\bm{ b_{2}\cdot x}}) + \cos(-({\bm{ b_{1}}}+{\bm{ b_{2}}) {\bm \cdot} {\bm x}})\big )$. Since the period of $W({\bm x})$ is described by the two reciprocal lattice vectors  $\bb_1$ and $\bb_2$, the non-zero density of states appears only at $\bp=\bk \pm \bb_1$, $\bk \pm \bb_2$, and $\bk \pm (\bb_1 + \bb_2)$.  As $\bk$ represents the momentum of the Dirac states, there are 12 duplications of the first order (the green and blue clones in Fig.~2{\bf e}) within the Brillouin zone of graphene, which is in agreement with the experimental observation. 

The last three terms of $\ket{\Psi_\bk}$ correspond to the second-order perturbations. The second last term on the right side of Eq.~(1) vanishes due to the fact that $\bra{\Psi_{\bm k}^0} W \ket{\Psi_{\bm k}^0}=0$. The last term only induces a renormalization of the primary cone at $\bk$. Only the first term of the second-order survives in certain conditions and give rise to clones in momentum space. To have non-vanishing $\bra{\Psi_{\bm l}^0} W \ket{\Psi_{\bm k}^0}$, the mediating momentum obeys ${\bm l}=\bk \pm \bb_1,\bk \pm \bb_2,\bk \pm \bb_1\pm\bb_2 $. Likewise, the momentum of the final wavefunction satisfies $\bp= \bk \pm 2 \bb_1, \bk \pm 2 \bb_2, \bk \pm 2 (\bb_1+ \bb_2), \bk\pm (\bb_1-\bb_2), \bk \pm (2\bb_1+\bb_2), \bk \pm (\bb_1+2\bb_2)$. In this regard, the second-order perturbations duplicate 24 Dirac cones (the gray and yellow clones in Fig.~2{\bf e}) at various $\bp$, consistent with  the experimental observation. 

The perturbative corrections to the wavefunction in Eq.~(1) give rise to the clones of Dirac cones. The clones represent a redistribution of spectral weight of the primary Dirac cone in the momentum space. That is why the observed clones share the same band dispersion as the primary Dirac cones. The clones derived from the same primary Dirac cone do not hybridize with each other. On the other hand, the hybridization are allowed for the clone or primary contours from different valleys. The hybridization between two valleys is mediated by the substrate potential. Here the two valleys of monolayer graphene behave like the two sets of Dirac cones from the two layers of TBG. Flat bands can be created under certain substrate conditions, which can be seen in the following discussions.

Our ARPES and tight-binding results indicate that the substrate potential places a periodic modulation to the graphene band structure, produces clones of Dirac states, and effectively shortens the distance between the two valleys by the reciporical vectors of the substrate. This machanism can enable a direct coupling between the Dirac states from the two valleys when the substrate lattice is nearly commensurate with graphene. Up to date, various graphene-based heterostructures such as  graphene/metals \cite{Rotenberg2015, HERNANDEZRODRIGUEZ201558, YU2019633}, graphene/boron nitride \cite{Yankowitz2019, Dean2010, Haigh2012}, and graphene/chalcogenide compounds \cite{grapheneVDW, Aeschlimanneaay0761, Mao2020, Geim2013, Ponomarenko2011, Georgiou2013} have been experimentally realized. To investigate the substrate effects in the nearly commensurate condition, we performed tight-binding simulation for a generic graphene heterostructure with a hexagonal substrate rotated by 30$^\circ$ relative to the graphene unit cell. The substrate lattice constant is chosen to be 3.8~\AA, which is about 10\% smaller than the commensurate value $\sqrt{3}a_\textrm{Gr} = 4.26\ \textrm{\AA}$. The cacluated band structure is shown in Fig.~4{\bf a}. At the Fermi level, there are two primary Dirac points (DP) denoted by D and D' and six duplicated DPs denoted by C1, C1', C2, C2', C3, and C3'. The clones of C1-C3 are from the valley of the `D' Dirac cone while those of C1'-C3' are from the other valley. When two Dirac bands from different valleys (for example, C1' and D, or C2 and C2') intersect, an energy gap is opened at the crossing point. The gapped band structure give rise to Van Hove singularities (VHS) in the density of states (DOS) as marked by the yellow and red arrows. The iso-energy contours at $E = -0.08$~eV is plotted in Fig.~4{\bf b}. Close to the zero energy, all the primary and cloned contours are isolated in momentum space and thus contribute to the DOS as independent Dirac cones. Therefore, the DOS vanishes at zero energy as shown in Fig.~4{\bf a}. The effective distance between the two primary Dirac points in the presence of substrate perturbations is 
\begin{equation}
q=\big||{\textrm K'}_\textrm{Gr} {\textrm K}_\textrm{Gr}|-b\big|=\big||\Gamma {\textrm K}_\textrm{Gr}|-b\big|, 
\end{equation}
where $b$ is the length of the substrate reciprocal lattice vector, as schematically shown in Fig.~4{\bf c}. That is also the separation between DPs C1' and D in Fig.~4{\bf b}. As the substrate constant approaches the commensurate value $\sqrt{3}a_\textrm{Gr}$, the clone contours move closer to the the primary cone and enhance the coupling between the two valleys, since the effective coupling is described by a dimensionless parameter $\alpha = \frac{w}{\hbar v_{\rm F} q}$, where $w$ is the amplitude of the substrate potential and $v_{\rm  F}$ is the Fermi velocity of electrons in graphene \cite{Bistritzer12233}. $w$ is set to be $0.05t =140$~meV (comparable to that of TBG, $w\approx110$~meV)\cite{Bistritzer12233}, where $t =2.8$~eV is the nearest-neighbor hopping parameter of graphene\cite{CastroNeto2009}. When the substrate lattice constant is equal to 4.166~\AA\ and 4.370~\AA, a sharp peak shows up at zero energy in DOS and an energy gap of size $\sim 2w$ emerges between the conduction and valence bands, as shown in Figs.~4{\bf e,f}. The zero-energy peaks in DOS cannot be described by isolated Dirac cones, therefore there must be dispersionless bands emerging at low energy as a consequence of hybridization of Dirac states from the two valleys.  

The effective separation between the two DPs of monolayer gaphene in the presence of substrate modulation is described by three vectors, ${\bm q_{1}}$ = $q$(1, 0),  ${\bm q_{2}}$ = $q$($\frac{-1}{2}$, $\frac{\sqrt{3}}{2}$), and ${\bm q_{3}}$ = $q$($\frac{-1}{2}$, $\frac{-\sqrt{3}}{2}$). Repeated hopping between the two valleys generates a $k$-space honeycomb lattice shown in Fig.~5{\bf a}. The unit vectors of this lattice are same as reciprocal vectors of the Moire pattern (MP) of this hybrid structure, namely, ${\bm b}^{\rm MP}_{1} = {\bm q_{1}}-{\bm q_{2}}$ and ${\bm b}^{\rm MP}_{2} = {\bm q_{1}}-{\bm q_{3}}$. The low-energy electron dynamics can be described by an effective Hamiltonian, 
\begin{equation}
H_{\rm eff}({\bm k})=\begin{bmatrix}
D({\bm k}) & T_{1} & T_{2} &  T_{3} \\
T^{\dagger}_1 & \overline{D}({\bm k-{\bm q_{1}}}) & 0 & 0 \\
T^{\dagger}_2 & 0 & \overline{D}({\bm k-{\bm q_{2}}}) & 0 \\
T^{\dagger}_3 & 0 & 0 & \overline{D}({\bm k-{\bm q_{3}}})
\end{bmatrix},
\end{equation}
where $D({\bm k})=\hbar v_{F}  {\bm k}{\bm \cdot} {\bm \sigma}$, $\overline{D}({\bm k})=-\hbar v_{F}  {\bm k}{\bm \cdot} {\bm \sigma^{*}}$, and $T_{m}=w\exp({i\frac{2(1-m)\pi}{3}}\sigma_{z})$, $m=$1, 2, and 3. Here ${\bm \sigma}=(\sigma_{x}, \sigma_{y}$) and $\sigma_{z}$ are Pauli matrices associated with the A and B sublattices of graphene. We note that $D({\bm k})$ and $\overline{D}({\bm k})$ describes the Dirac cone at $\rm K$ and $\rm K'$, respectively, as the two cones are time-reversal partners. The derivation of the effective Hamiltonian can be found in the supplementary information. Using this effective Hamiltonian, we calculated the band structure along the path A-B-C-D-A and the DOS, see Fig.~5{\bf b}. For $\alpha$ = 0.1, the DPs at B and C remained isolated meanwhile an energy gap is opened at the crossing point of two Dirac bands. The DOS shows a nearly linear dependence of energy and several peaks from VHSs, which is consistent with the tight-binding result in Fig.~4{\bf a}. When $\alpha$ increases to 0.586, a pair of absolutely flat bands exist at zero energy inside the band gap ($\Delta E \sim 2 w$). This leads to a sharp zero-energy peak in DOS, which agrees remarkably well with the tight-binding DOS with $a_{\rm sub}$ = 4.370~\AA\ and $w=0.05t$ as shown in the inset of Fig.~5{\bf b}. The perfect flatness of zero-energy bands is due to the chiral symmetry of the Hamiltonian, since $H_{\rm eff}$ is equivalent to the chiral-symmetric continuum model proposed by Tarnopolsky {\it et al.}\cite{PhysRevLett.122.106405} (see the proof in the supplementary information). The intervalley transition matrices $T_m$ ($m$ = 1, 2, and 3) in $H_{\rm eff}$ contain only the diagonal AA and BB couplings due to the on-site substrate potential and the fact that the wavefunctions of Dirac states at K and K' are defined with respect to the same A and B sublattices of monolayer graphene. The absence of AB couplings yields the chiral symmetry of the effective Hamiltonian and creates the absolutely flat bands. In other words, this system is a natural realization of the chiral-symmetric model of flat bands\cite{PhysRevLett.122.106405}. The unique coupling $\alpha^{*}=0.586$ corresponds to two `magic' lattice constants according to Eq.~2, 
\begin{equation}
a^{*}_{\rm sub}=\sqrt{3}a_{\rm Gr}\pm \frac{3wa_{\rm Gr}}{2\pi t\alpha^{*}} .
\end{equation}
Plug in $\alpha^{*}=0.586$, $a_{\rm Gr}$ = 2.46 \AA\ and $w=0.05t$, we find that $a^{*}_{\rm sub}$ = 4.161 and  4.361 \AA. The magic lattice constants echo the values (4.166 and 4.370 \AA) we found in the tight-binding simulations. The small discrepancy between two results can be attributed to the finite size of the supercell we used in the tight-binding simulations. The effective Hamiltonian gives rise to a series of magic coupling $a^{*}$ with a periodicity of $\Delta\alpha \simeq 1.5$ \cite {Bistritzer12233, PhysRevLett.126.016404}. The second magic coupling is $a^{*}$ = 2.221. The band structure with this magic coupling (the bottom panel of Fig.~5{\bf b}) exhibits absolute flat bands at zero energy and a smaller band gap ($\Delta E \sim 0.2w$). We notice that other bands such as the ones at $\pm$0.25$w$ are also flattened, leading sharp spikes in the DOS. The second and higher magic couplings correspond to a very small deviation from the commensurate lattice constant, $\Delta a \lesssim 0.02$\ \AA. It is technically challenging to detect such small lattice deviations in experiments, just like the smaller twists corresponding the higher magic couplings in the TBG systems. 

The low-energy dynamics in the graphene heterostructures is essentially governed by the coupling $\alpha$, and $\alpha$ is determined by the separation $q$ between two adjacent Dirac cones in the Moir{\' e} lattice. For $w=0.05t$ ($\sim$140~meV) and $\alpha^{*}=0.586$, $q$ = 0.04\ ${\rm \AA}^{-1} \ll |\Gamma{\rm K}_{\rm Gr}| = 1.7\ {\rm \AA}^{-1}$, which means the reciprocal lattice vectors of substrate must almost connect the two DPs at K and K'  in momentum space, see Fig.~4{\bf c}. This places a constraint on the possible substrate lattice constants and orientations. To have flat bands, the largest possible substrate lattice constant corresponds to a $(\sqrt{3}\times\sqrt{3})R30^{\circ}$ supercell of graphene, $i.e.$, $a_{\rm sub}\approx\sqrt{3}a_{\rm Gr}=4.26$~ \AA. This commensurate relation is also known as the Kekul{\'e} superlattice\cite{PhysRevB.88.155415}. For other commensurate relations between graphene and substrates, the substrate lattice constant has to be $\lesssim \frac{\sqrt{3}}{2}a_{\rm Gr}=2.13$~\AA, which is very rare in real materials. Therefore, the substrate materials for the flat-band heterostructure must have a surface with $C_3$ rotation symmetry and lattice constant close to 4.26~\AA. In Table~1, we list several materials which can be potentially employed in the proposed heterostructures. In addition to the conventional assembly method by growing or transferring graphene onto the substrate surface, the proposed flat-band heterostructures can be readily synthesized via a “top-down” approach, as schematically plotted in Fig.~5{\bf c}. High-quality graphene layers can be epitxially grown on the SiC(0001) surface, and the Dirac states of graphene remain isolated due to the large lattice mismatch. Therefore, the graphene/SiC structure can serve as a supporting substrate for the growth of various materials with nearly commensurate relations as suggested in Table~1. For example, Bi$_{2}$Se$_{3}$ ($a$ = 4.136~\AA) thin layers have been grown on the graphene/SiC(0001) surface \cite{Jin2013}. 

In conclusion, our ARPES experiments demonstrated the cloning of Dirac fermions in the graphene/SiC(0001) heterostructure due to the periodic modulation of the substrate potential. Our theoretical calculations showed this modulation effect from the substrate can effectively couple the two valleys of Dirac states in monolayer graphene in the nearly commensurate condition. The graphene heterostructures can be a promising alternate system for exploring the intriguing flat-band physics that was found in TBG. The criterion for realizing flat bands is a matchup of the surface potential strength and the periodicity of the substrate to reach the magic effective coupling $\alpha^{*}$. There are a vast selections of substrate materials that can be potentially used in this hybrid structure. In addition, the charge and spin orderings in the substrates such as antiferromagnetism in MnTe\cite{PhysRevB.96.214418}, superconductivity in PdTe$_2$\cite{PhysRevB.97.014523}, and topological surface states in Bi$_{2}$Se$_{3}$\cite{Jin2013} can further enrich flat-band physics in graphene via proximity effects, systematic investigations of which is left to future works.

\begin{table}[ht]
\centering 
\begin{tabular}{c c c c c c} 
\hline\hline 
Materials &\ $a_{\rm sub}$ (\AA) &\ Space group &\ \ \ \ \ \ \ \ \  \ \ \ \ \ \ \ \ \ Materials &\ $a_{\rm sub}$ (\AA) &\ Space group \\ [0.5ex] 
\hline 
CuSe & 3.980 & $P6_{3}/mmc$ &\ \ \ \ \ \ \ \ \   \ \ \ \ \ \ \ \ \ Bi$_{2}$Se$_{3}$ & 4.136 & $R$-$3mH$ \\
InSe	& 4.000 & $R$-$3mH$	&\ \ \ \ \ \ \ \ \   \ \ \ \ \ \ \ \ \ CdS & 4.137 & $P6_{3}/mc$ \\
CrTe	& 4.005 & $P6_{3}/mmc$ &\ \ \ \ \ \ \ \ \   \ \ \ \ \ \ \ \ \ MnTe & 4.148 & $P6_{3}/mmc$ \\
PdTe$_{2}$ & 4.024 & $P$-$3m1$ &\ \ \ \ \ \ \ \ \   \ \ \ \ \ \ \ \ \ 	GeTe & 4.156 & $R$-$3mH$ \\
In$_{2}$Se$_{3}$ & 4.026 & $R$-$3mH$ &\ \ \ \ \ \ \ \ \   \ \ \ \ \ \ \ \ \ PdTe & 4.200 & $P6_{3}/mmc$ \\
PtTe$_{2}$ & 4.026 & $P$-$3m1$	 &\ \ \ \ \ \ \ \ \   \ \ \ \ \ \ \ \ \ CdSe & 4.232 & $P6_{3}/mc$ \\
InSe & 4.050 & $P6_{3}/mmc$ &\ \ \ \ \ \ \ \ \   \ \ \ \ \ \ \ \ \ Cu$_2$Te & 4.237 & $P6/mmm$ \\
As$_2$Te$_3$ & 4.058 & $R3mH$ &\ \ \ \ \ \ \ \ \   \ \ \ \ \ \ \ \ \ Sb$_2$Te$_3$ & 4.264 & $R$-$3mH$ \\
GaTe & 4.060 & $P6_{3}/mmc$ &\ \ \ \ \ \ \ \ \   \ \ \ \ \ \ \ \ \ SiTe$_2$ & 4.289 & $P$-$3m1$ \\
ZnTe	 & 4.092 & $P3_{1}21$ &\ \ \ \ \ \ \ \ \   \ \ \ \ \ \ \ \ \ MgSe & 4.319 & $P6_{3}/mmc$ \\
ScTe	 & 4.097 & $P6_{3}/mmc$ &\ \ \ \ \ \ \ \ \   \ \ \ \ \ \ \ \ \ HgSe & 4.320	 & $P3_{2}21$ \\
AuTe$_2$ & 4.107 & $P$-$3m1$ &\ \ \ \ \ \ \ \ \   \ \ \ \ \ \ \ \ \ Sb(111) & 4.332 & $R$-$3mH$ \\
PtTe & 4.111 & $P6_{3}/mmc$ &\ \ \ \ \ \ \ \ \   \ \ \ \ \ \ \ \ \ HgTe & 4.392 & $P3_{1}21$ \\
AuSe & 4.120 & $P6_{3}/mmc$ &\ \ \ \ \ \ \ \ \   \ \ \ \ \ \ \ \ \ Bi$_2$Te$_3$ & 4.403 & $R$-$3mH$ \\
MnSe & 4.120 & $P6_{3}/mc$ &\ \ \ \ \ \ \ \ \   \ \ \ \ \ \ \ \ \ MgTe & 4.531 & $P6_{3}/mmc$ \\
Cu$_2$Se & 4.132 & $R$-$3mH$ &\ \ \ \ \ \ \ \ \   \ \ \ \ \ \ \ \ \ Bi(111) & 4.546 & $R$-$3mH$ \\
\hline 
\end{tabular}
\caption{{\bf Material candidates for flat-band heterostructures.} The materials are ordered according to their in-plane lattice constant. Bi(111) and Sb(111) are Bi and Sb films grown in the rhombohedral (111) direction\cite{BIAN2019109}.} 
\end{table}

\section{Methods}
\subsection{Tight-bind modeling}
To simulate the observed spectrum of the graphene on the top of the SiC film, we use the simplest graphene model with the inclusion of only the nearest neighbor hopping. Since SiC is an insulator with large gaps, the main low energy physics can be described by the graphene monolayer embedded in the electric potential on the SiC surface. Although the details of the SiC surface potential is unknown, it suffices to employ an approximate potential that preserves the crystalline symmetry. The SiC surface belongs to the wallpaper group $P3m1$, thus the SiC potential can be written in this simplest form 
\begin{align}
W(x,y)=&w\big(\cos (\pi \frac{4y}{3b}) +\cos (\pi \frac{-2\sqrt{3}x-2y}{3b}) + \cos (\pi \frac{2\sqrt{3}x-2y}{3b})\big )
\end{align}
where $b$ (1.77~\AA) is the in-plane silicon-carbon distance on the SiC(0001) surface and $\bb_{1,2}= \frac{2\pi}{3b} (\pm \sqrt{3},-1)$. That is, the SiC potential includes only the first order of the Fourier series and breaks the $C_6$ rotation symmetry due to the different potentials stemming from Si and C atoms. We can write the effective low-energy Hamiltonian in the second quantization form 
\begin{align}
\hat{H}=&\sum_{n,m}\Big [t( a^\dagger_{n,m }b_{n,m}+ a^\dagger_{n+1,m}b_{n,m}+a^\dagger_{n,m+1}b_{n,m}+{\rm{h.c.}}) + W(x_{nm}^+,y_{nm}^+)a^\dagger_{n,m }a_{n,m} \nonumber\\
&+W(x_{nm}^-,y_{nm}^-)b^\dagger_{n,m }b_{n,m} \Big ],
\end{align}
where $x_{nm}^\pm = (\mp 1 + 3 n -3 m)a/2,\ y_{nm}^\pm  = (\sqrt{3} n +\sqrt{3} m)a/2$ and $a\approx 1.42$~\AA\ is the carbon-carbon distance in graphene. A 400$\times$400 supercell is employed in the calculations of band structure and DOS. By diagonalizing the Hamiltonian, we find the eigenstates within $E\pm\delta $. Then we transform the eigenstates to momentum space so that at energy $E$, the density of the wavefunctions can be plotted in momentum space. 

\subsection{Sample synthesis and characterizations}
The graphene films were prepared by annealing a 6H-SiC(0001) substrate at 1150 °C in an integrated MBE-STM-ARPES ultrahigh vacuum (UHV) system with base pressure below 2$\times$10$^{-10}$~mbar. After the growth, the graphene samples were {\it in-situ} transferred the ARPES stage. ARPES measurements were performed at 100 K using a SPECS~PHOIBOS-150 hemisphere analyzer with a SPECS UVS-300 helium discharge lamp (photon energy = 21.2 eV). The size of the beam spot on the sample was $\sim$1.5~mm. The topography of the sample surface was mapped {\it in-situ} by an Aarhus STM equipped in the growth chamber.

The TEM samples were prepared by a lift-out method in a ThermoFisher Scientific Scios focused ion beam (FIB) instrument at room temperature, and imaged in the ThermoFisher Scientific G2 Tecnai F30 FEG high resolution TEM operated at 300 kV. The SiC substrate was tilted to the [100] zone axis and the lattice fringes from both the graphene and the SiC can be clearly resolved. Great care has been taken to reduce the beam damage on the thin film samples both during the FIB lift out and during the sample tilting and high-resolution image acquisition process.

\newpage
\begin{figure}
\centering
\includegraphics[width=16cm]{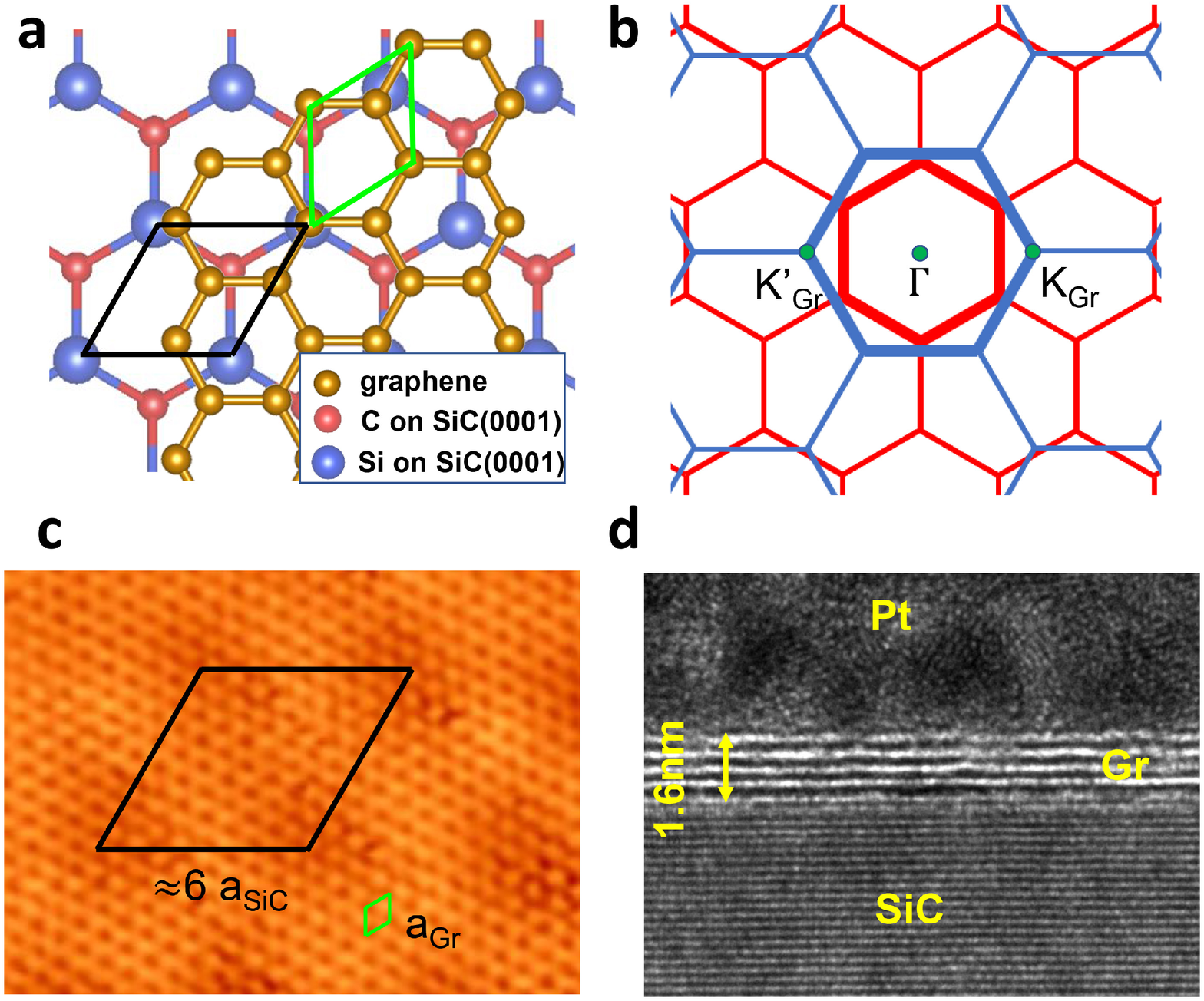}
\caption{{\bf Lattice structure and Brillouin zone of graphene/SiC heterostructure.} {\bf a,} Lattice structure of graphene and SiC(0001) surface. The unit cell of graphene and SiC surface is depicted by the green and black parallelograms, respectively. {\bf b,} Brillouin zone of graphene (blue lines) and SiC(0001) surface (red lines). {\bf c,} STM image of graphene with atomic resolution. A Moir\'{e} pattern is observed with a period approximately equal to 6 times the lattice constant of SiC(0001). {\bf d,} Cross-sectional TEM image of graphene/SiC heterostructure.}
\end{figure}

\newpage
\begin{figure}
\centering
\includegraphics[width=16cm]{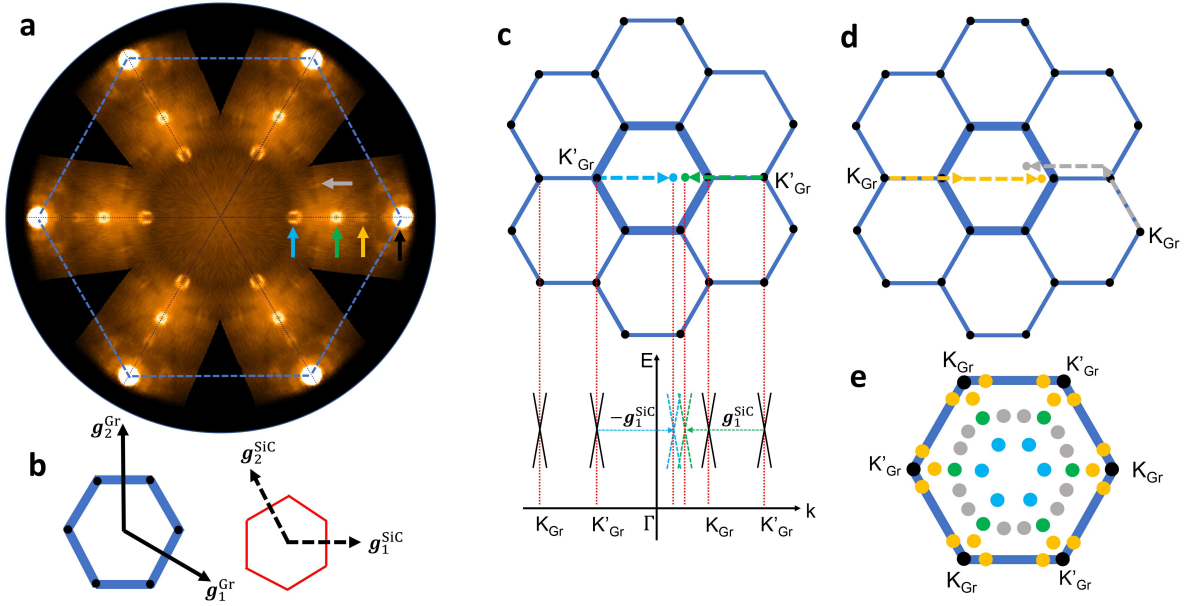}
\caption{{\bf ARPES spectrum and momentum-space analysis of Dirac fermion cloning.}  {\bf a,} Fermi surface of graphene/SiC(0001) heterostructure measured by ARPES.  {\bf b,} Geometrical relation between the Brillouin zones of graphene and SiC(0001) surface.  {\bf c,} The location of green and blue clones from the first-order perturbations.  {\bf d,} The location of yellow and gray clones from the second order perturbations.  {\bf e,} Distribution of first-order and second-order clones in the Brillouin zone of graphene.}
\end{figure}

\newpage
\begin{figure}
\centering
\includegraphics[width=16cm]{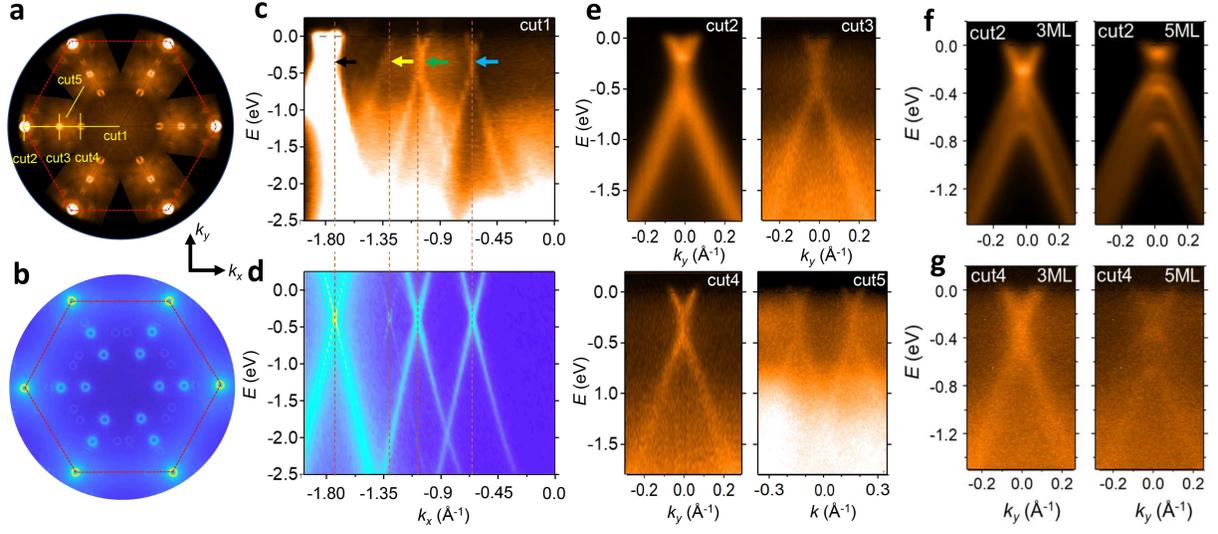}
\caption{{\bf ARPES spectra and tight-binding simulation of Dirac bands and their clones.} {\bf a,}  ARPES and {\bf b,}  tight-binding fermi surface of graphene/SiC heterostructure. {\bf c,}  ARPES and {\bf d,}  tight-binding spectrum taken along `cut1' marked in {\bf a}. {\bf e,}  ARPES spectra taken along `cut2-5'.   {\bf f,}  ARPES spectrum of the primary Dirac cone taken along `cut2' from 3-, 5-monolayer graphene films grown on SiC(0001) surface. {\bf g,}  ARPES spectrum of cloned Dirac bands taken along `cut4' from 3-, 5-monolayer graphene films.}
\end{figure}

\newpage
\begin{figure}
\centering
\includegraphics[width=16cm]{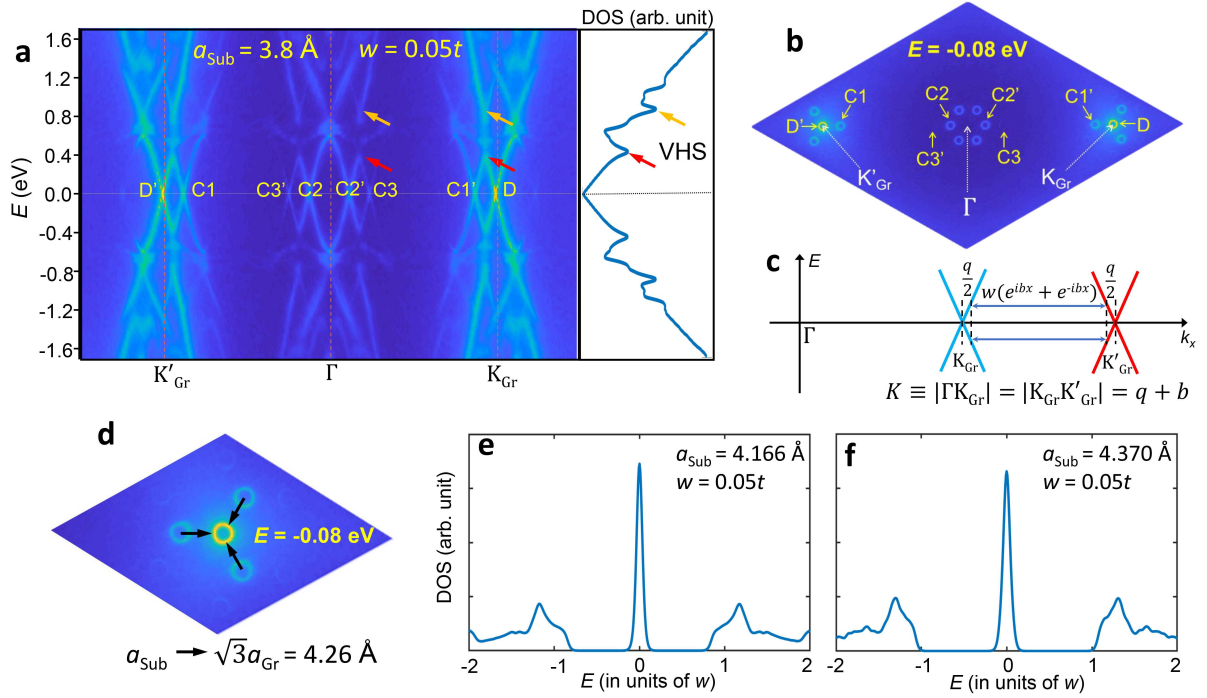}
\caption{{\bf Tight-binding simulation of epitaxial graphene with various substrate lattice constants.} {\bf a,} The tight-binding band structure and density of states with $a_\textrm{sub} = 3.8\ \textrm{\AA}$. {\bf b,}  Calculated iso-energy contours at $E = -0.08\ eV$. {\bf c,} Coupling mechanism between the two Dirac cones at ${\textrm K_\textrm{Gr}}$ and ${\textrm K'_\textrm{Gr}}$. {\bf d,}  Schematic of the movement of clone contours C1'  as the substrate approaches to the commensurate value $\sqrt{3}a_\textrm{Gr}= 4.26\ \textrm{\AA}$. {\bf e,} Calculated density of states for $a_\textrm{sub} = 4.166\ \textrm{\AA}$ and $w = 0.05t$. {\bf f,} Same as {\bf e} but for $a_\textrm{sub} = 4.370\ \textrm{\AA}$.}
\end{figure}

\newpage
\begin{figure}
\centering
\includegraphics[width=16cm]{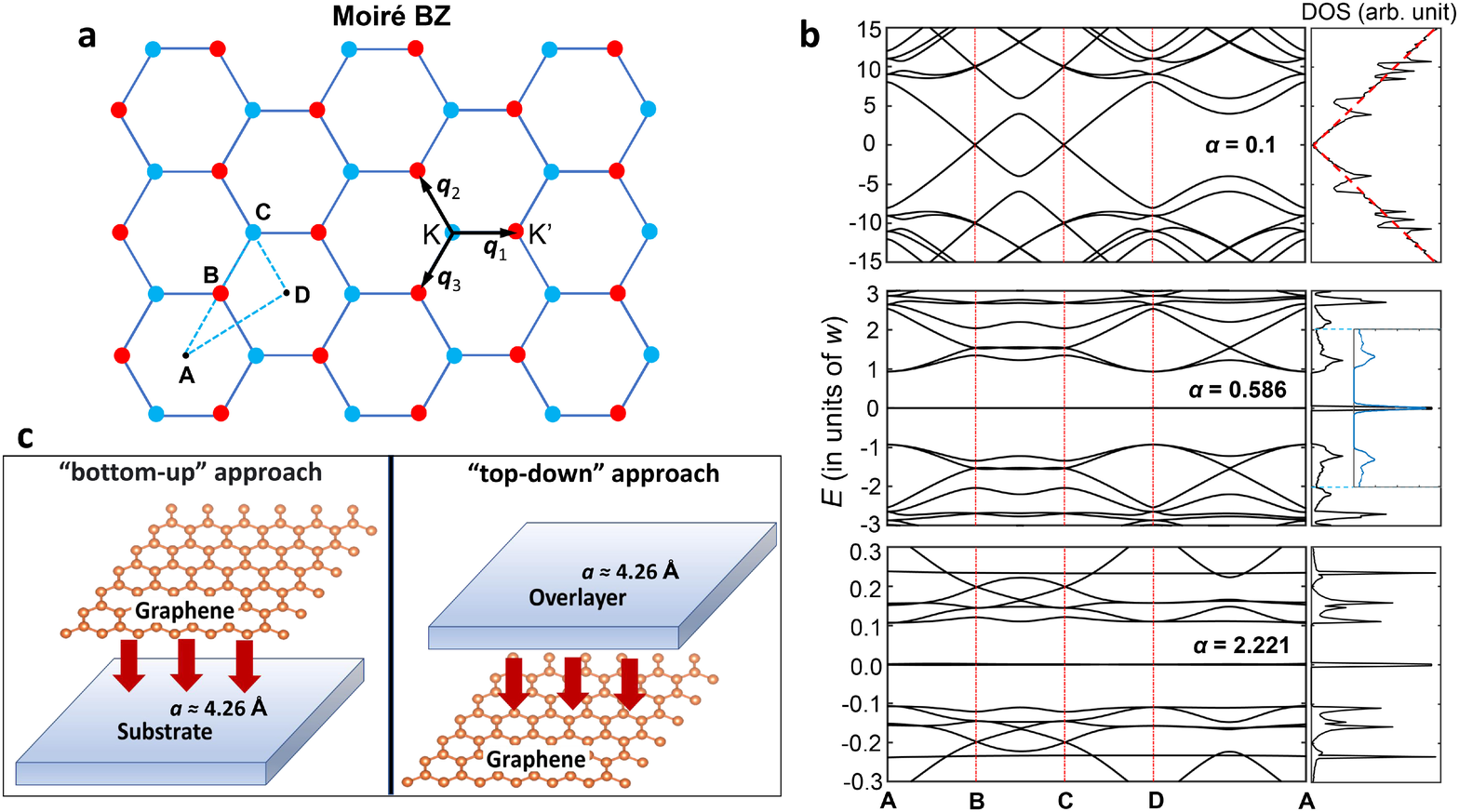}
\caption{{\bf Effective model of Moir{\'e} mini lattice and flat bands.}
{\bf a,} Moir{\'e} Brillouin zone of a nearly commensurate graphene heterostructure. {\bf b,} Calculated band structure and density of states with $\alpha = 0.1, 0.586$, and $2.221$. The DOS plot in the middle panel includes an inset showing the tight-binding DOS (the blue curve) from Fig.~4{\bf f}. {\bf c,} Schematic of the two ways of assembling heterostructures with flat bands.}
\end{figure}

\bibliographystyle{naturemag}
\bibliography{graphene}

\begin{thebibliography}{10}
\expandafter\ifx\csname url\endcsname\relax
  \def\url#1{\texttt{#1}}\fi
\expandafter\ifx\csname urlprefix\endcsname\relax\def\urlprefix{URL }\fi
\providecommand{\bibinfo}[2]{#2}
\providecommand{\eprint}[2][]{\url{#2}}

\bibitem{CastroNeto2009}
\bibinfo{author}{{Castro Neto}, A.~H.}, \bibinfo{author}{Guinea, F.},
  \bibinfo{author}{Peres, N. M.~R.}, \bibinfo{author}{Novoselov, K.~S.} \&
  \bibinfo{author}{Geim, A.~K.}
\newblock \bibinfo{title}{{The electronic properties of graphene}}.
\newblock \emph{\bibinfo{journal}{Reviews of Modern Physics}}
  \textbf{\bibinfo{volume}{81}}, \bibinfo{pages}{109--162}
  (\bibinfo{year}{2009}).

\bibitem{Novoselov2005}
\bibinfo{author}{Novoselov, K.~S.} \emph{et~al.}
\newblock \bibinfo{title}{{Two-dimensional gas of massless Dirac fermions in
  graphene}}.
\newblock \emph{\bibinfo{journal}{Nature}} \textbf{\bibinfo{volume}{438}},
  \bibinfo{pages}{197--200} (\bibinfo{year}{2005}).

\bibitem{Zhang2005}
\bibinfo{author}{Zhang, Y.}, \bibinfo{author}{Tan, Y.-W.},
  \bibinfo{author}{Stormer, H.~L.} \& \bibinfo{author}{Kim, P.}
\newblock \bibinfo{title}{{Experimental observation of the quantum Hall effect
  and Berry's phase in graphene}}.
\newblock \emph{\bibinfo{journal}{Nature}} \textbf{\bibinfo{volume}{438}},
  \bibinfo{pages}{201--204} (\bibinfo{year}{2005}).

\bibitem{Geim2013}
\bibinfo{author}{Geim, A.~K.} \& \bibinfo{author}{Grigorieva, I.~V.}
\newblock \bibinfo{title}{{Van der Waals heterostructures}}.
\newblock \emph{\bibinfo{journal}{Nature}} \textbf{\bibinfo{volume}{499}},
  \bibinfo{pages}{419--425} (\bibinfo{year}{2013}).

\bibitem{Katsnelson2006}
\bibinfo{author}{Katsnelson, M.~I.}, \bibinfo{author}{Novoselov, K.~S.} \&
  \bibinfo{author}{Geim, A.~K.}
\newblock \bibinfo{title}{{Chiral tunnelling and the Klein paradox in
  graphene}}.
\newblock \emph{\bibinfo{journal}{Nature Physics}}
  \textbf{\bibinfo{volume}{2}}, \bibinfo{pages}{620--625}
  (\bibinfo{year}{2006}).

\bibitem{RevModPhys.86.959}
\bibinfo{author}{Basov, D.~N.}, \bibinfo{author}{Fogler, M.~M.},
  \bibinfo{author}{Lanzara, A.}, \bibinfo{author}{Wang, F.} \&
  \bibinfo{author}{Zhang, Y.}
\newblock \bibinfo{title}{Colloquium: Graphene spectroscopy}.
\newblock \emph{\bibinfo{journal}{Rev. Mod. Phys.}}
  \textbf{\bibinfo{volume}{86}}, \bibinfo{pages}{959--994}
  (\bibinfo{year}{2014}).

\bibitem{PhysRevLett.61.2015}
\bibinfo{author}{Haldane, F. D.~M.}
\newblock \bibinfo{title}{Model for a quantum hall effect without landau
  levels: Condensed-matter realization of the "parity anomaly"}.
\newblock \emph{\bibinfo{journal}{Phys. Rev. Lett.}}
  \textbf{\bibinfo{volume}{61}}, \bibinfo{pages}{2015--2018}
  (\bibinfo{year}{1988}).

\bibitem{RevModPhys.82.3045}
\bibinfo{author}{Hasan, M.~Z.} \& \bibinfo{author}{Kane, C.~L.}
\newblock \bibinfo{title}{Colloquium: Topological insulators}.
\newblock \emph{\bibinfo{journal}{Rev. Mod. Phys.}}
  \textbf{\bibinfo{volume}{82}}, \bibinfo{pages}{3045--3067}
  (\bibinfo{year}{2010}).

\bibitem{RevModPhys.83.1057}
\bibinfo{author}{Qi, X.-L.} \& \bibinfo{author}{Zhang, S.-C.}
\newblock \bibinfo{title}{Topological insulators and superconductors}.
\newblock \emph{\bibinfo{journal}{Rev. Mod. Phys.}}
  \textbf{\bibinfo{volume}{83}}, \bibinfo{pages}{1057--1110}
  (\bibinfo{year}{2011}).

\bibitem{Liu2011}
\bibinfo{author}{Liu, Y.}, \bibinfo{author}{Bian, G.}, \bibinfo{author}{Miller,
  T.} \& \bibinfo{author}{Chiang, T.-C.}
\newblock \bibinfo{title}{{Visualizing Electronic Chirality and Berry Phases in
  Graphene Systems Using Photoemission with Circularly Polarized Light}}.
\newblock \emph{\bibinfo{journal}{Physical Review Letters}}
  \textbf{\bibinfo{volume}{107}}, \bibinfo{pages}{166803}
  (\bibinfo{year}{2011}).

\bibitem{Bistritzer12233}
\bibinfo{author}{Bistritzer, R.} \& \bibinfo{author}{MacDonald, A.~H.}
\newblock \bibinfo{title}{Moir{\'e} bands in twisted double-layer graphene}.
\newblock \emph{\bibinfo{journal}{Proceedings of the National Academy of
  Sciences}} \textbf{\bibinfo{volume}{108}}, \bibinfo{pages}{12233--12237}
  (\bibinfo{year}{2011}).

\bibitem{Ponomarenko2013}
\bibinfo{author}{Ponomarenko, L.~A.} \emph{et~al.}
\newblock \bibinfo{title}{{Cloning of Dirac fermions in graphene
  superlattices}}.
\newblock \emph{\bibinfo{journal}{Nature}} \textbf{\bibinfo{volume}{497}},
  \bibinfo{pages}{594--597} (\bibinfo{year}{2013}).

\bibitem{Dean2013}
\bibinfo{author}{Dean, C.~R.} \emph{et~al.}
\newblock \bibinfo{title}{{Hofstadter's butterfly and the fractal quantum Hall
  effect in moir{\'{e}} superlattices}}.
\newblock \emph{\bibinfo{journal}{Nature}} \textbf{\bibinfo{volume}{497}},
  \bibinfo{pages}{598--602} (\bibinfo{year}{2013}).

\bibitem{Cao20181}
\bibinfo{author}{Cao, Y.} \emph{et~al.}
\newblock \bibinfo{title}{{Correlated insulator behaviour at half-filling in
  magic-angle graphene superlattices}}.
\newblock \emph{\bibinfo{journal}{Nature}} \textbf{\bibinfo{volume}{556}},
  \bibinfo{pages}{80--84} (\bibinfo{year}{2018}).

\bibitem{Cao20182}
\bibinfo{author}{Cao, Y.} \emph{et~al.}
\newblock \bibinfo{title}{{Unconventional superconductivity in magic-angle
  graphene superlattices}}.
\newblock \emph{\bibinfo{journal}{Nature}} \textbf{\bibinfo{volume}{556}},
  \bibinfo{pages}{43--50} (\bibinfo{year}{2018}).

\bibitem{Wang2016}
\bibinfo{author}{Wang, E.} \emph{et~al.}
\newblock \bibinfo{title}{{Gaps induced by inversion symmetry breaking and
  second-generation Dirac cones in graphene/hexagonal boron nitride}}.
\newblock \emph{\bibinfo{journal}{Nature Physics}}
  \textbf{\bibinfo{volume}{12}}, \bibinfo{pages}{1111--1115}
  (\bibinfo{year}{2016}).

\bibitem{Sharpe605}
\bibinfo{author}{Sharpe, A.~L.} \emph{et~al.}
\newblock \bibinfo{title}{Emergent ferromagnetism near three-quarters filling
  in twisted bilayer graphene}.
\newblock \emph{\bibinfo{journal}{Science}} \textbf{\bibinfo{volume}{365}},
  \bibinfo{pages}{605--608} (\bibinfo{year}{2019}).

\bibitem{Serlin900}
\bibinfo{author}{Serlin, M.} \emph{et~al.}
\newblock \bibinfo{title}{Intrinsic quantized anomalous hall effect in a
  moir{\'e} heterostructure}.
\newblock \emph{\bibinfo{journal}{Science}} \textbf{\bibinfo{volume}{367}},
  \bibinfo{pages}{900--903} (\bibinfo{year}{2020}).

\bibitem{Kaxiras2020}
\bibinfo{author}{Tritsaris, G.~A.} \emph{et~al.}
\newblock \bibinfo{title}{Electronic structure calculations of twisted
  multi-layer graphene superlattices}.
\newblock \emph{\bibinfo{journal}{2D Materials}} \textbf{\bibinfo{volume}{7}},
  \bibinfo{pages}{035028} (\bibinfo{year}{2020}).

\bibitem{PhysRevLett.122.106405}
\bibinfo{author}{Tarnopolsky, G.}, \bibinfo{author}{Kruchkov, A.~J.} \&
  \bibinfo{author}{Vishwanath, A.}
\newblock \bibinfo{title}{Origin of magic angles in twisted bilayer graphene}.
\newblock \emph{\bibinfo{journal}{Phys. Rev. Lett.}}
  \textbf{\bibinfo{volume}{122}}, \bibinfo{pages}{106405}
  (\bibinfo{year}{2019}).

\bibitem{Conrad2017}
\bibinfo{author}{Conrad, M.} \emph{et~al.}
\newblock \bibinfo{title}{{Wide band gap semiconductor from a hidden 2D
  incommensurate graphene phase}}.
\newblock \emph{\bibinfo{journal}{Nano Letters}} \textbf{\bibinfo{volume}{17}},
  \bibinfo{pages}{341--347} (\bibinfo{year}{2017}).

\bibitem{PhysRevB.96.035411}
\bibinfo{author}{Huang, L.} \emph{et~al.}
\newblock \bibinfo{title}{Effects of moir\'e lattice structure on electronic
  properties of graphene}.
\newblock \emph{\bibinfo{journal}{Phys. Rev. B}} \textbf{\bibinfo{volume}{96}},
  \bibinfo{pages}{035411} (\bibinfo{year}{2017}).

\bibitem{Zhou2006}
\bibinfo{author}{Zhou, S.~Y.} \emph{et~al.}
\newblock \bibinfo{title}{{First direct observation of Dirac fermions in
  graphite}}.
\newblock \emph{\bibinfo{journal}{Nature Physics}}
  \textbf{\bibinfo{volume}{2}}, \bibinfo{pages}{595--599}
  (\bibinfo{year}{2006}).

\bibitem{Liu2010}
\bibinfo{author}{Liu, Y.} \emph{et~al.}
\newblock \bibinfo{title}{{Phonon-induced gaps in graphene and graphite
  observed by angle-resolved photoemission}}.
\newblock \emph{\bibinfo{journal}{Physical Review Letters}}
  \textbf{\bibinfo{volume}{105}}, \bibinfo{pages}{1--4} (\bibinfo{year}{2010}).

\bibitem{Rotenberg2015}
\bibinfo{author}{Rotenberg, E.} \& \bibinfo{author}{Bostwick, A.}
\newblock \bibinfo{title}{{Superlattice effects in graphene on SiC(0001) and
  Ir(111) probed by ARPES}}.
\newblock \emph{\bibinfo{journal}{Synthetic Metals}}
  \textbf{\bibinfo{volume}{210}}, \bibinfo{pages}{85--94}
  (\bibinfo{year}{2015}).

\bibitem{HERNANDEZRODRIGUEZ201558}
\bibinfo{author}{Hernández-Rodríguez, I.}, \bibinfo{author}{García, J.~M.},
  \bibinfo{author}{Martín-Gago, J.~A.}, \bibinfo{author}{{de Andrés}, P.~L.}
  \& \bibinfo{author}{Méndez, J.}
\newblock \bibinfo{title}{Graphene growth on pt(111) and au(111) using a mbe
  carbon solid-source}.
\newblock \emph{\bibinfo{journal}{Diamond and Related Materials}}
  \textbf{\bibinfo{volume}{57}}, \bibinfo{pages}{58 -- 62}
  (\bibinfo{year}{2015}).

\bibitem{YU2019633}
\bibinfo{author}{Yu, J.} \emph{et~al.}
\newblock \bibinfo{title}{Study on aln buffer layer for gan on graphene/copper
  sheet grown by mbe at low growth temperature}.
\newblock \emph{\bibinfo{journal}{Journal of Alloys and Compounds}}
  \textbf{\bibinfo{volume}{783}}, \bibinfo{pages}{633 -- 642}
  (\bibinfo{year}{2019}).

\bibitem{Yankowitz2019}
\bibinfo{author}{Yankowitz, M.}, \bibinfo{author}{Ma, Q.},
  \bibinfo{author}{Jarillo-Herrero, P.} \& \bibinfo{author}{LeRoy, B.~J.}
\newblock \bibinfo{title}{{van der Waals heterostructures combining graphene
  and hexagonal boron nitride}}.
\newblock \emph{\bibinfo{journal}{Nature Reviews Physics}}
  \textbf{\bibinfo{volume}{1}}, \bibinfo{pages}{112--125}
  (\bibinfo{year}{2019}).

\bibitem{Dean2010}
\bibinfo{author}{Dean, C.~R.} \emph{et~al.}
\newblock \bibinfo{title}{{Boron nitride substrates for high-quality graphene
  electronics}}.
\newblock \emph{\bibinfo{journal}{Nature Nanotechnology}}
  \textbf{\bibinfo{volume}{5}}, \bibinfo{pages}{722--726}
  (\bibinfo{year}{2010}).

\bibitem{Haigh2012}
\bibinfo{author}{Haigh, S.~J.} \emph{et~al.}
\newblock \bibinfo{title}{{Cross-sectional imaging of individual layers and
  buried interfaces of graphene-based heterostructures and superlattices}}.
\newblock \emph{\bibinfo{journal}{Nature Materials}}
  \textbf{\bibinfo{volume}{11}}, \bibinfo{pages}{764--767}
  (\bibinfo{year}{2012}).

\bibitem{grapheneVDW}
\bibinfo{author}{Zhang, Z.} \emph{et~al.}
\newblock \bibinfo{title}{Graphene-based mixed-dimensional van der waals
  heterostructures for advanced optoelectronics}.
\newblock \emph{\bibinfo{journal}{Advanced Materials}}
  \textbf{\bibinfo{volume}{31}}, \bibinfo{pages}{1806411}
  (\bibinfo{year}{2019}).

\bibitem{Aeschlimanneaay0761}
\bibinfo{author}{Aeschlimann, S.} \emph{et~al.}
\newblock \bibinfo{title}{Direct evidence for efficient ultrafast charge
  separation in epitaxial ws2/graphene heterostructures}.
\newblock \emph{\bibinfo{journal}{Science Advances}}
  \textbf{\bibinfo{volume}{6}} (\bibinfo{year}{2020}).

\bibitem{Mao2020}
\bibinfo{author}{Mao, J.} \emph{et~al.}
\newblock \bibinfo{title}{{Evidence of flat bands and correlated states in
  buckled graphene superlattices}}.
\newblock \emph{\bibinfo{journal}{Nature}} \textbf{\bibinfo{volume}{584}},
  \bibinfo{pages}{215--220} (\bibinfo{year}{2020}).

\bibitem{Ponomarenko2011}
\bibinfo{author}{Ponomarenko, L.~A.} \emph{et~al.}
\newblock \bibinfo{title}{{Tunable metal–insulator transition in double-layer
  graphene heterostructures}}.
\newblock \emph{\bibinfo{journal}{Nature Physics}}
  \textbf{\bibinfo{volume}{7}}, \bibinfo{pages}{958--961}
  (\bibinfo{year}{2011}).

\bibitem{Georgiou2013}
\bibinfo{author}{Georgiou, T.} \emph{et~al.}
\newblock \bibinfo{title}{{Vertical field-effect transistor based on
  graphene–WS2 heterostructures for flexible and transparent electronics}}.
\newblock \emph{\bibinfo{journal}{Nature Nanotechnology}}
  \textbf{\bibinfo{volume}{8}}, \bibinfo{pages}{100--103}
  (\bibinfo{year}{2013}).

\bibitem{PhysRevLett.126.016404}
\bibinfo{author}{Ren, Y.}, \bibinfo{author}{Gao, Q.},
  \bibinfo{author}{MacDonald, A.~H.} \& \bibinfo{author}{Niu, Q.}
\newblock \bibinfo{title}{Wkb estimate of bilayer graphene's magic twist
  angles}.
\newblock \emph{\bibinfo{journal}{Phys. Rev. Lett.}}
  \textbf{\bibinfo{volume}{126}}, \bibinfo{pages}{016404}
  (\bibinfo{year}{2021}).

\bibitem{PhysRevB.88.155415}
\bibinfo{author}{Wallbank, J.~R.},
  \bibinfo{author}{Mucha-Kruczy\ifmmode~\acute{n}\else \'{n}\fi{}ski, M.} \&
  \bibinfo{author}{Fal'ko, V.~I.}
\newblock \bibinfo{title}{Moir\'e minibands in graphene heterostructures with
  almost commensurate $\sqrt{3}\ifmmode\times\else\texttimes\fi{}\sqrt{3}$
  hexagonal crystals}.
\newblock \emph{\bibinfo{journal}{Phys. Rev. B}} \textbf{\bibinfo{volume}{88}},
  \bibinfo{pages}{155415} (\bibinfo{year}{2013}).

\bibitem{Jin2013}
\bibinfo{author}{Jin, K.-H.} \& \bibinfo{author}{Jhi, S.-H.}
\newblock \bibinfo{title}{{Proximity-induced giant spin-orbit interaction in
  epitaxial graphene on a topological insulator}}.
\newblock \emph{\bibinfo{journal}{Physical Review B}}
  \textbf{\bibinfo{volume}{87}}, \bibinfo{pages}{075442}
  (\bibinfo{year}{2013}).

\bibitem{PhysRevB.96.214418}
\bibinfo{author}{Kriegner, D.} \emph{et~al.}
\newblock \bibinfo{title}{Magnetic anisotropy in antiferromagnetic hexagonal
  mnte}.
\newblock \emph{\bibinfo{journal}{Phys. Rev. B}} \textbf{\bibinfo{volume}{96}},
  \bibinfo{pages}{214418} (\bibinfo{year}{2017}).

\bibitem{PhysRevB.97.014523}
\bibinfo{author}{Das, S.} \emph{et~al.}
\newblock \bibinfo{title}{Conventional superconductivity in the type-ii dirac
  semimetal ${\mathrm{pdte}}_{2}$}.
\newblock \emph{\bibinfo{journal}{Phys. Rev. B}} \textbf{\bibinfo{volume}{97}},
  \bibinfo{pages}{014523} (\bibinfo{year}{2018}).

\bibitem{BIAN2019109}
\bibinfo{author}{Bian, G.} \emph{et~al.}
\newblock \bibinfo{title}{Survey of electronic structure of bi and sb thin
  films by first-principles calculations and photoemission measurements}.
\newblock \emph{\bibinfo{journal}{Journal of Physics and Chemistry of Solids}}
  \textbf{\bibinfo{volume}{128}}, \bibinfo{pages}{109--117}
  (\bibinfo{year}{2019}).

\end{thebibliography}

\end{document}